\documentclass[12pt]{article}
\usepackage{amsfonts,amssymb,amsmath,amscd}
\usepackage{epsf}

\def\Pf{{\rm Pf}}
\def\Fhat{\hat{F}}

\def\({\left(}
\def\){\right)}
\def\[{\left[}
\def\]{\right]}

\def\({\left(}
\def\){\right)}

\def\Fhat{\hat{F}}

\def\Pf{{\rm Pf}} 
\def\Fhat{\hat{F}}

\def\beq{\begin{equation}}
\def\eeq{\end{equation}}
\def\bea{\begin{eqnarray}}
\def\eea{\end{eqnarray}}
\def\bq{\begin{quote}}
\def\eq{\end{quote}}

\def\Pf{{\rm Pf}} 
\def\Fhat{\hat{F}}

\def\({\left(}
\def\){\right)}

\def\g5{\gamma_5}
\def\gappeq{\mathrel{\rlap {\raise.5ex\hbox{$>$}}
{\lower.5ex\hbox{$\sim$}}}}

\def\lappeq{\mathrel{\rlap{\raise.5ex\hbox{$<$}}
{\lower.5ex\hbox{$\sim$}}}}

\def\Toprel#1\over#2{\mathrel{\mathop{#2}\limits^{#1}}}

\begin{document}

\pagestyle{empty}
\begin{flushright}
{CPHT-RR 005.0204}\\ 
{\ttfamily hep-th/0403126}\\ 
\end{flushright}
\vspace*{5mm}
\begin{center}
{ \Huge Branes as Stable Holomorphic Line Bundles
 \\
 \vspace*{3mm} \Huge  On the Non-Commutative
Torus }\\
\vspace*{19mm}
{\large Pascal Grange} \\
\vspace{0.5cm}

 {\it Centre de physique th{\'e}orique de l'{\'E}cole polytechnique,\\
 \vspace*{0.2cm}route de Saclay, 91128 Palaiseau Cedex, France}\\
\vspace{0.5cm} 
{\tt{ pascal.grange@cpht.polytechnique.fr}}\\
\vspace*{3cm}

\vspace*{1.5cm}  
{\bf ABSTRACT} \\ \end{center}
\vspace*{5mm}
\noindent

 It was recently suggested by A. Kapustin that turning on a $B$-field, and
 allowing some discrepancy between the left and and right-moving complex
 structures, must induce an identification of B-branes with
 holomorphic line bundles on a 
 non-commutative complex torus. We translate the stability condition
 for the branes into this language and identify the stable topological
 branes with previously proposed non-commutative instanton
 equations. This involves certain topological identities whose
 derivation has become familiar in non-commutative field theory. It is
 crucial for these identities that the instantons are localized. We
 therefore  explore the case of non-constant field strength,
  whose non-linearities are dealt with thanks to the rank-one Seiberg--Witten map.

\vspace*{0.5cm}
\noindent

\begin{flushleft} 
February 2004 
\end{flushleft}

\vfill\eject

\setcounter{page}{1}
\pagestyle{plain}

\section{Introduction}

Branes of various dimensions may be regarded as submanifolds of a target
space, since their position is defined by a set of boundary
conditions for open strings. As open strings carry gauge fields,
taking those into account promotes branes to bundles on the
 submanifolds. Background fields from closed strings influence the geometry of these bundles; in particular,
{\mbox{non-commutative}} gauge theory becomes a valid tool for the field theory
along the branes, when a $B$-field is turned on. Many couplings between open and closed
string modes follow from the relevant star-products~\cite{CH,SW,Cornalba,DMS}. Branes also allow for formulations of important
field-theoretic issues, such as the problem
of instantons. Investigating supersymmetric D$p$-branes through
an effective action, Mari{\~n}o,
Minasian, Moore and Strominger~\cite{MMMS} identified a class of
instanton equations, and proposed a {\mbox{non-commutative}} limit thereof,
\begin{equation}\label{holomorphicity}  \hat{F}^{(0,2)}=0,\end{equation}
\begin{equation}\label{stability}\hat{F}\wedge J^{\frac{p-1}{2}}=0.\end{equation}
 On the other hand, Kapustin~\cite{Kapustin} recently used the tools of (generalized)
 complex geometry~\cite{Hitchin,Gualtieri} to identify the influence of the $B$-field on
  topological branes. He argued that it endows {\mbox{B-branes}} on the
 {\mbox{non-commutative}} torus with a structure of holomorphic
  line bundles, the complex structure of which is dictated by the
 $B$-field and
  an allowed discrepancy between left and {\mbox{right-moving}} complex
  structures. Without {\mbox using} a {\mbox{non-commutative}}
 description, {\mbox{Kapustin}}
  and Li
  also obtained~\cite{KapustinWS} a stability condition, in agreement
  with~\cite{MMMS}, using   
  the world-sheet viewpoint, which makes branes emerge as boundary
 conditions that are compatible with a supersymmetry algebra.\\

The purpose of the present note is two-fold: I shall put together
these two viewpoints,  and perform
checks required by the non-linearities of {\mbox{non-commutative}} gauge
theory with {\mbox{non-constant}} field strength.  I shall adapt the stability condition to the
{\mbox{non-commutative}} {\mbox{set-up}}, thereby showing that the {\mbox{non-commutative}}
instanton equations
of~\cite{MMMS} are equivalent to the identification, in the presence
of a $B$-field, between
supersymmetric D-branes and stable holomorphic line bundles on
the {\mbox{non-commutative}} torus. But both formulations were written down on
the basis of explicit formulae in the case of constant field
strength; their natural extension to {\mbox{non-constant}} ones, involving
highly non-linear terms in the gauge field, remains conjectural. The
checks will  use the techniques of {\mbox{non-commutative}} gauge theory, in order to deal with
some of these non-linearities, and to establish the {\mbox{non-commutative}}
instanton equations, that were originally written down without relying on
these techniques.\\

The structure of this paper is as follows. In section 2, I shall review
the geometric set-up in which the {\mbox{non-commutative}} proposal for holomorphicity was formulated,
and translate into this language the stability condition of matching of spectral flow
operators. In section 3, I shall explain how the holomorphicity
condition~(\ref{holomorphicity})  cancels terms that would not be consistent
with the Seiberg--Witten map in flat space. The constant factor
appearing in the stability condition will finally be used to produce a
topological identity, whose Seiberg--Witten limit reads as the equation~(\ref{stability}).

\section{A class of stable line bundles}
\subsection{Topological branes and generalized complex geometry}
Consider a (B)-brane as a line bundle over a complex
 torus. According to a proposal in~\cite{Kapustin}, turning on a $B$-field, and
 allowing a discrepancy between the left and and right-moving complex
 structures $I_+$ and $I_-$, can give rise to holomorphic line bundles over a
 {\mbox{non-commutative}} complex torus. The way to this proposal goes as
 follows. If $X$ is a torus, endowed with a metric $G$ and a
 $B$-field, a pair of complex structures $I_+$ and $I_-$ can be used
 to define a complex structure on the sum $X\oplus X^\ast$, suited to
 the even-dimensional cases, where $X$ is the quotient of a complex
 vector space by a lattice,
$$
\begin{pmatrix}
I_+ & 0\\
 0 & I_-
\end{pmatrix}.\\
$$

 The $B$-field allows for transformations of complex structures on
$X\oplus X^\ast$ discussed in~\cite{Hitchin, Gualtieri} and
related to isometries of $X\oplus X^\ast$. On the other hand, the
presence of a metric yields an isomorphism between tangent and
cotangent spaces, and $I_+$
and $I_-$ can be interpreted as complex structures for left and
right-movers respectively. Given the fact that T-duality is a parity
transformation acting on right-movers only, the most general data (involving
different left and right-moving complex structures) are relevant to the
study of T-duality. It will be assumed that the following
 complex structure on the sum of tangent and cotangent spaces 
$$
\mathcal{I}=\begin{pmatrix}\tilde{I} +(\delta P)B & -\delta P \\
\delta\omega+B(\delta P)B+B\tilde{I} +\tilde{I}^tB & -\tilde{I}^t-B\delta P
\end{pmatrix},\\
$$
is block-upper-triangular. This assumption is related to the
transformation by T-duality of a block-diagonal complex structure with
no discrepancy between left and right-moving complex structures. It
allows us to investigate how {\mbox{non-commutative}} deformations and
discrepancies between complex structures are entangled.  The
tensor $\tilde{I}$ is half of the sum of the
left and {\mbox{right-moving}} complex structures; the discrepancies $\delta
P$ and $\delta\omega$ are defined through the difference between the
associated Kaehler forms $\omega_{\pm}=G I_\pm$ through
$$\delta\omega=\frac{1}{2}(\omega_+-\omega_-),$$
$$\delta P=\frac{1}{2}(\omega_+ ^{-1}-\omega_- ^{-1}),$$
 Boundary conditions for fermions and supersymmetry requirements
have been shown~\cite{Kapustin} to give rise under this assumption to the
following equation for the field strength,
\begin{equation}\label{default}FI+I ^tF=-F\delta P F,\end{equation}
so that the failure of the field strength to be of type $(1,1)$ is
 directly related to the difference allowed between left and
 right-moving complex structures. This failure has furthermore been related
  to non-commutativity by proposing that, given the two relations 
$$I=\tilde{I} +(\delta P)B,$$
$$\delta P=I\theta+\theta I ^t,$$
the failure disappears, provided one considers
 the basis as a torus with non-commutativity scale
 $\theta= B^{-1}$, and the field strength $\hat{F}$ associated  to $F$
 by the Seiberg--Witten map in flat space:
\begin{equation}
\label{proposal}\hat{F} I+ I ^t\hat{F}=0.
\end{equation}

 This proposal means that the {\mbox{non-commutative}} counterpart of the
field strength is of type $(1,1)$, and thereby ensures that $N=2$
supersymmetry compells the brane to be a holomorphic line bundle on
a {\mbox{non-commutative}} complex torus.  A few consistency checks were
performed in~\cite{Kapustin} for constant field strength, by substituting
$\hat{F}=(1+\theta F)^{-1} F$ in the proposal~(\ref{proposal}) and expanding it in
powers of $\theta F$. In the present note, I shall adopt the reverse
approach, substituting the inverse Seiberg-Witten map
in~(\ref{default}), and identifying the condition~(\ref{proposal}) as the
suitable tool to make the expansion consistent. Apart from checking
the proposal more precisely, the consideration of varying field
strength will prove necessary for the consistency with the stability
condition for the bundle, because this condition involves the
assumption of localized instantons. Instanton equations derived from
{\mbox{non-commutative}} field theory and supersymmetric branes
in~\cite{SW,MMMS} rely indeed on the existence of a region with the
combination $B+F$ merely consisting of a constant $B$-field.

\subsection{Stability condition}

In order to make contact with (some limit of) the deformed equations proposed
in~\cite{MMMS} for supersymmetric D-branes, we have to supplement the
previous proposal with a stability requirement, because stable
topological branes correspond to supersymmetric D-branes. To this end,
we are going to rephrase, using the language of left and right-moving
complex
 structures, the stability condition derived by Kapustin and
 Li~\cite{KapustinWS} using world-sheet arguments. Their work 
  recovers the condition
 of~\cite{MMMS} by considering the holomorphic part of the matching
 condition for spectral flows.  As the
complex structure $I$ in the previous statement has been unambiguously
identified, making it  appear as the complex structure suited to the
stability condition will amount to a consistency check.\\

 In terms of
a holomorphic $n$-form, where $n$ is the complex dimension of the
ambient manifold, the matching condition for the spectral flows reads
in terms of a proportionality factor $e ^{i\alpha}$ as follows:
$$\Omega_{i_1\dots \i_n}\psi_+ ^{i_1}\dots\psi_+
^{i_n}=e ^{i\alpha}\,\Omega_{i_1\dots \i_n}\psi_- ^{i_1}\dots\psi_- ^{i_n}.$$
In order to make contact with the above discussion, we first have to
rewrite the latter equation terms of the linear combinations 
$$\psi ^i=\frac{1}{2}\(\psi_+^i +\psi_-^i\) ,$$
$$\rho_i=\frac{1}{2} G_{ij}\(\psi_+^j -\psi_-^j\),$$
for which the boundary conditions read 
$$\rho_i=-(B_{ij}+F_{ij}) \psi ^j.$$
Complex structures induce a splitting of boundary conditions. We
can consider the holomorphic part, whose determinant was expressed
in~\cite{KapustinWS} using the discussion in terms of the variables $(\psi_+,
\psi_-)$. If the boundary conditions for these variables are written
using some linear transformation $R$, the stability condition for
B-branes is worked out
by identifying the constant factor $e ^{i\alpha}$ with the determinant
of the holomorphic part $R_h$ of the transformation:
$$\psi_+=R\psi_-,$$
$$\det R_h= e ^{i\alpha}.$$

 Let us adapt this statement to the variables $(\psi,\rho)$.  
As the stability requirement is to be eventualy expressed in terms of a
Kaehler form, it will involve the holomorphic part of the boundary
conditions, and therefore a complex structure. Under the assumption of
 $\mathcal{I}$ being block-upper-triangular, we have to check that $\rho$
 and $\psi$ make for a $(1,0)$-form on the direct sum of the tangent
 and cotangent spaces, according to a complex structure on the sum of
 tangent and cotangent spaces  written in the basis adapted to
 $(\psi,\rho)$.
The matching condition written in terms of $\psi$ is still
independent from complex structures, and is expressed in terms of the
combination $\mathcal{F}:=B+F$, as:
$$\det\( G+\mathcal{F}\)= e ^{i\alpha}\det\(G- \mathcal{F}\).$$
 But once the boundary condition has been
imposed, linking $\psi$ and $\rho$ to each other, it must
be compatible with the consideration of the
holomorphic part of the transformation of the tangent bundle. This
means that, if  $\psi$ is holomorphic with respect to $I$, the
fermion $\rho$ associated to $\psi$ by  the boundary condition must be such that
$(\psi,\rho)$ is holomorphic with respect to the
complex structure 
$$
\begin{pmatrix}
I_+ & 0\\
 0 & I_-
\end{pmatrix},\\  
$$
once written in the new
basis that is adapted to sections  of tangent and cotangent spaces
such as $(\psi, \rho)$, namely
$$
\begin{pmatrix}
\tilde{I} & -\delta P \\ 
\delta\omega & -\tilde{I}^t
\end{pmatrix}.\\
$$
 This is seen to be guaranteed by the assumption of
$\mathcal{I}$ being block-upper-triangular, since the holomorphicity condition on
$\psi ^i$ reads from the blocks on the first line, together with the
{\mbox{boundary}} conditions:
$$\tilde{I}\psi +\(\delta P\) B \psi=i\psi ,$$
and is sufficient to ensure the relation corresponding to the blocks
on the second line, since the latter reads:
$$\(-B(\delta P) B-B\tilde {I} -\tilde{I} ^t B\)\psi +\tilde{I}^t
B\psi=-i B\psi.$$
The formulation we need for the stability condition is thus in
terms of the Kaehler form
$$J:= G \(\tilde{I}+\(\delta P\)B\)=GI,$$
which is suited to the consideration of the holomorphic part of the matching
condition. In
order to evaluate the factor in terms of our fields, we use the gauge-freedom exchange
between $B$ and $F$ to express stability through 
$$\frac{\det(J+i\mathcal{F})}{\det(1+G
    ^{-1}\mathcal{F})}=\frac{\det(J+iB)}{\det(1+G ^{-1}B)},$$
were we assumed that the instanton is localized. We
 evaluated the constant factor far away from the instanton, where
 $\mathcal{F}$ reduces to the $B$-field. As was announced, the field
 strength cannot be constant, and the
 identification between branes and stable holomorphic line bundles
 therefore requires a treatment of the non-linearities carried by the
 Seiberg--Witten map of {\mbox{non-commutative}} gauge theory.

\section{Compatibility with the Seiberg--Witten map}
\subsection{Holomorphic line bundles and the
  Seiberg--Witten map} 
I shall now perform a few checks of the proposal~(\ref{proposal}) for varying field
strength. These computations are motivated by the stability condition,
as just stated, but will be performed on the holomorphicity condition,
thus preparing for the combination of these two as an
instanton equation.

In order to obtain constraints on the {\mbox{non-commutative}} gauge theory from
 the constraint~(\ref{default}) on the field strength, we must formulate this condition in terms of
 {\mbox{non-commutative}} fields in position space. The rank-one
 Seiberg--Witten map in flat space~\cite{MW, Liu, OO} admits a natural formulation in
  momentum space, since it involves a straight open Wilson line $W_k$
 whose extension $\theta ^{\mu\nu}k _\nu$ depends on the Fourier mode $k_\mu$ in consideration:
\begin{equation} \label{map}F_{ij}(k)=\int dx\, L_\ast\left(\sqrt{\det(1-\theta
    \hat{F})},\,\hat{F}_{ik}\(\frac{1}{1-\theta\hat{F}}\)^k_j,\, W_k(x)\right) \ast e
    ^{ikx},\end{equation}
where $L_\ast$ denotes the smearing prescription: it averages over all
the  possible ways of inserting operators along the Wilson
line. Expanding such a smeared expression in powers of the gauge field
defines the modified $\ast_n$ products, where the integer $n$ labels
    the order of the expansion. The {\sc{LHS}} of the equation
$$FI+I ^tF=-F\delta P F$$ 
only contains one operator in position space. We may thus 
rewrite it, using the Seiberg--Witten map, as the Fourier transform of the expression
$$\int dx\, L_\ast\left(\sqrt{\det(1-\theta
    \hat{F})},\left(\hat{F}\frac{1}{1-\theta\hat{F}}I+ I ^t\hat{F}\frac{1}{1-\theta\hat{F}}\right),W_k(x)\right) \ast e
    ^{ikx},$$
since objects carrying no index are transparent to the endomorphism
$I$. The {\sc{RHS}} is more involved, since it consists of a pointwise product
of two fields of gauge theory in position space. Its Fourier transform
at momentum $k_\mu$ will therefore be a convolution of two Wilson lines,
whose extensions sum to $\theta^{\mu\nu}k_\nu$, each of these Wilson
lines having one of the fields attached at its beginning. This is
equivalent to the concatenation of two Wilson lines into one, of extension
$\theta^{\mu\nu}k_\nu$, the second observable being smeared along the
 result. For momentum $k_\mu$, we obtain in momentum space 
$$\int dx\, L_\ast\left(\sqrt{\det(1-\theta
    \hat{F})},\hat{F}\frac{1}{1-\theta\hat{F}},\left(I\theta\hat{F}\frac{1}{1-\theta\hat{F}}+ \theta I ^t\hat{F}\frac{1}{1-\theta\hat{F}}\right),W_k(x)\right) \ast e
    ^{ikx}.$$
Since differential operators $\ast_n$ enable us to expand both
expressions at all orders in derivatives and any finite order in
the gauge field, we can read off the conditions for the expansions of
the two sides of~(\ref{default}) to
match order by order in the gauge field, even for varying fields.\\ 

 The first order in the field strength is consistent with 
 $\hat{F}$ being of type $(1,1)$:
$$\hat{F}I+I ^t\hat{F}=o(\hat{F}).$$
  The next order makes appear a single term
 on the {\sc{RHS}}, because each of the form indices has to be brought by a
 field strength, with no contribution either from the determinant or
 from the denominators. It has a counterpart on the {\sc{LHS}}, through the
 quadratic terms in the following expansion of the Seiberg--Witten map in
 position space:  
$$F_{ij}=\Fhat_{ij} +\theta ^{mn} \langle\Fhat_{im},\Fhat_{nj}\rangle_{\ast_2}
-\frac{1}{2}\theta ^{mn} \langle\Fhat_{nm},\Fhat_{ij}\rangle_{\ast_2} +\theta ^{mn}
\partial_n\langle \hat{A}_m, \Fhat_{ij}\rangle_{\ast_2} + O(\Fhat^3).$$
 Now we can read off the condition to be fulfilled in order for the
 two sides to match at 
 quadratic order in the gauge field:
$$0= -\frac{1}{2} \theta ^{mn}\langle\Fhat_{nm},(\Fhat I + I ^t\Fhat)_{ij}\rangle_{\ast_2} +\theta ^{mn}
\partial_n\langle \hat{A}_m,(\Fhat I + I ^t\Fhat)_{ij}\rangle_{\ast_2}.$$
We observe that it is identically verified if the holomorphicity condition obtained at
the linear step is fulfilled. This is the first test of the
non-commutative holomorphicity proposal passed by varying fields.\\

Expanding further in terms of the gauge field will
generate infinitely many more involved contributions, but we may note at once
that the condition $\hat{F}^{(0,2)}=0$ will ensure the cancellation of
an infinite subset, which can only be obtained on the {\sc{LHS}}, namely the
subset of those terms with form indices borne by one field strength,
and with an arbitrary number of gauge fields from the expansion of the
open Wilson line:
$$\sum_{p\geq 1}\frac{i ^p}{p!}(\theta\partial)^{\lambda_1}\dots(\theta\partial)^{\lambda_p}\left\langle\sqrt{\det\(1-\theta\hat{F}\)},\hat{F}_{ij},\hat{A}_{\lambda_1},\dots,
\hat{A}_{\lambda_p}\right\rangle_{\ast_p}.$$
 
 We have altogether seen that the condition $\hat{F}^{(0,2)}=0$ is
exactly what we need to cancel a set of terms (of arbitrarily high
degree in the gauge field) that does only appear on
the {\sc{LHS}}. In order to achieve a more convincing argument, we have to
explain inductively why the remaining
terms match. Since the influence of the Pfaffian and the Wilson line
is the same on both sides, we just have to deal with the terms coming
from the expansion of the denominator in the Seiberg--Witten map~(\ref{map}). These correspond to terms with the
two form-indices borne by two different field strengths, one being the
field strength in the numerator, the other one being the last factor
in a term of the series expansion of the denominator. Going from
some definite order in the expansion in powers of the gauge field to
the next one involves a variation with respect to the discrepancy
$\delta P$, because the latter is quadratic in the field strength. But the form
of the terms thereby generated is precisely the one that comes from
expansions of denominators. Consider a certain order $p$ in the
expansion of $\hat{F}(1-\theta\hat{F})^{-1}$ in powers of the field
strength inside the {\sc{LHS}} of~(\ref{default}), and suppose we have been able to show that the terms of
interest match up to that order. Consider the effect of making the discrepancy act
on one of the field strengths, say the $A$-th, thus producing, up to a sign, a term
with one more field strength. We have to sum over all the ways of
making this insertion, so that the monomial of order $p+1$ in the
field strength to be smeared on the {\sc{LHS}} reads 
$$\sum_{A=1}^p(\theta\hat{F})^{A-1} \hat{F}\(\delta P\)\hat{F}(\theta \hat{F})^{p-A}.$$
Rewriting its smearing as a $\ast_{p+1}$, we recognize a term that is
produced by taking $A$ powers of $\hat{F}$ from the first operator and
$p+1-A$ from the second one in the expansion of
$$L_\ast\(\frac{\hat{F}}{(1-\theta\hat{F})}, (\delta
P)\frac{\hat{F}}{(1-\theta\hat{F})}, W_k \),$$
which coincides with the list of operators smeared on the {\sc{RHS}}, when
the Pfaffian is disregarded.
 Restoring the contributions from 
the Pfaffian and the Wilson line on both sides ensures consistency between the two expansions
of the {\mbox{non-commutative}} image of~(\ref{default}), up to terms
that are cancelled on a holomorphic line bundle through~(\ref{proposal}).\\

\subsection{Stability condition as a {\mbox{non-commutative}} topological identity}

The derivation in the previous section relied on the various operators
that arise when agreement is demanded between commutative and {\mbox{non-commutative}}
couplings to {\mbox{Ramond--Ramond}} fields, in the case of a single D$p$-brane. The Seiberg--Witten map embodies
this agreement for couplings to $C^{p-1}$, and more field strengths are
inserted along the Wilson line when couplings to {\mbox{Ramond--Ramond}} fields of lower
degree are written. Nevertheless, the coupling to the top-form
$C^{p+1}$ yields one more identity, although no field strength can
carry form-indices in this coupling.  
 For a flat brane, the commutative coupling is a zero form that does not depend
 on the field strength. On the other hand, the {\mbox{non-commutative}}
 expression still has to be gauge-invariant, and to involve additional
 gauge 
 fields, even if all indices are contracted. This
 provides~\cite{DMS,CSterms} an
 identity between a gauge-invariant {\mbox{non-commutative}} expression, and a
 commutative expression that is actually more than
 gauge-invariant. The commutative side does not know about the
 variations of the {\mbox{non-commutative}} field strength, hence the name {\it{topological
    identity}}. It comes out as the zero-form
part{\footnote{Integration is along the $p+1$-dimensional
    world-volume, and one only retains the couplings in which the sum
    of form-degrees from Ramond--Ramond fields and gauge fields equals $p+1$.}} of the identity
between Ramond--Ramond couplings
$$\sum_n C^{(n)}(-k) \wedge\int dx\, \(e ^{B+F} \,e ^{ikx}\)=\sum_n
C^{(n)}(-k)\wedge\int dx \,L_{\ast}\( \frac{\rm{Pf} Q}{\rm{Pf} \theta},e
^{Q ^{-1}}, W_k(x)\)\ast e ^{ikx},$$
where $Q$ is the (inverse of the) {\mbox{non-commutative}} counterpart of the
symplectic structure $B+F$, in the case of constant field strength:
$$ Q^{ij}=\(\frac{1}{B+F}\)^{ij}=\theta ^{ij}-\theta^{ik}
\hat{F}_{kl}\theta ^{lj}.$$
The topological part is the following, whereas higher-degree
contributions yield the Seiberg-Witten map used in the previous
checks, and derivative corrections: 
$$\delta(k)=\int
dx\;L_\ast\(\sqrt{\det\(1-\theta\hat{F}\)},W_k(x)\)\ast e ^{ikx}.$$ 

This situation, where a gauge-invariant quantity is actually a
constant, is reminiscent of the presence of a constant factor in
the requirement of stability. The biais introduced by the complex structure forces us to adapt the
 above procedure to  the case at hand, endowed  with complex
 geometry. Let us note that this way of  
 producing identities, using the duality between commutative and
 {\mbox{non-commutative}} descriptions, has been proven by explicit
 computations to be sensible~\cite{DMS,MukhiSP,mezigue}. We are allowed to repeat it in our
 context as soon as we can
 go through the following steps:\\ 
{\it 1. consider a gauge-invariant quantity depending on
 commutative gauge fields and $B$-fields,\\
2.  write it in terms of
 {\mbox{non-commutative}} variables in the case of constant field strength,\\
3. remember that the result must be gauge-invariant, and restore
 gauge-invariance by the {\mbox{smearing}} prescription.}\\
In the cases where the
 quantity is simply a constant, the procedure generates a topological
 identity.\\

 In the case at hand, the obvious candidate for the
 gauge-invariant quantity is the constant factor $e ^{i\alpha}$.
 If $J$ is the Kaehler form associated to the complex structure $I$,
 our topological quantity is expressed as
$$e ^{i\alpha}=\frac{\det( J+i Q^{-1}) }{ \det( G+ Q^{-1})},$$
which for constant field strength can be expressed in two
different ways, using either the symplectic structure
$B$ (away from the instanton), or the inverse of the tensor $Q^{ij}$
defined in {\mbox{non-commutative}} gauge
theory.
 
Of course, this simple substitution of different  expressions for the
same tensor 
will suffer from a lack of gauge-invariance in the case of
 varying field strength. This is a general problem related to the
 non-locality of {\mbox{non-commutative}}
 gauge theory, solved by the smearing prescription along an open Wilson
 line~\cite{Rey2,Gross,DT}.  
The {\mbox{non-commutative}} expression for constant field strength leads to
the following identity in the Seiberg--Witten limit 
$$ \frac{\(Q^{-1}\)^{\frac{p-1}{2}}\wedge J}{\Pf \(Q^{-1}\)} =
\frac{B^{\frac{p-1}{2}}\wedge J}{\Pf (B)},$$
which is exactly the limit worked out in~\cite{MMMS}, where 
 the identity between the two expressions for $Q$ was used to write
 down the instanton equations in the {\mbox{non-commutative}}
 set-up. These {\mbox{non-commutative}} conditions are equivalent to $\hat{F}$
 being of type $(1,1)$, together with the condition  
$$\hat{F}\wedge J^{\frac{p-1}{2}}=0,$$
whose gauge-invariant completion is just the analogous smeared
expression. This completes a world-sheet derivation
of~(\ref{holomorphicity}) and~(\ref{stability}). Their space-time
derivation in~\cite{MMMS} was formal and relied on the explicit
Seiberg--Witten map for constant field strength, although the result
has obvious extension to more general configurations corresponding to
localized instantons. We have justified this extension by mapping the
bundles to {\mbox{non-commutative}} gauge theory, using further properties of
star-products, that actually define the Seiberg--Witten image of
instantons.

\section{Conclusions}
We have put together a proposed {\mbox{non-commutative}} version of  the
instanton equations
obtained from supersymmetric D-branes using effective actions, and the
 stable holomorphic line bundles that emerge from the study of
 topological branes on the {\mbox{non-commutative}} torus. We thereby established more
 firmly the relevance of {\mbox{non-commutative}} gauge theory for the study of
  instantons. On the way we identifed one of
 the equations as a topological identity of {\mbox{non-commutative}} gauge
 theory, associated to the existence of a constant factor in the
 stability condition. Complex structures fit into
 techniques involved in generating topological indentities and
 derivative corrections to effective actions from {\mbox{non-commutative}}
 field theory. This was to be awaited since an alternative world-sheet derivation
 should exist for the results based on the study of branes as
 supersymmetric solitons.\\ 

 Nevertheless, the full instanton equations derived in~\cite{MMMS} are a
 two-parameter deformation of usual instanton equations. The limit we
 investigated corresponds to the Seiberg--Witten limit of
 {\mbox{non-commutative}} gauge theory. It is a very peculiar case, just as the
 derivative corrections to effective actions derived from
 non-commutativity in the Seiberg--Witten limit are a subset of those
 that can be derived by going beyond this limit, further deforming the
 star-product. In a sense, dealing with the
 non-linearities as we did amounts to taking into account a more
 precise effective action, involving $\ast_n$-products between fields
 entering couplings of rank $2n$. Recovering the two-parameter
 deformation should involve the deformations $\tilde{\ast}_n$ that
 appeared in~\cite{MSbeyond,beyond}. Very
 roughly, we can note that the first contribution from the metric to
 these  products is
 quadratic, just as in the deformed instanton equations of~\cite{MMMS}.\\  

 The issue of the inclusion of scalars in stability conditions has been investigated~\cite{variations}, but exhibiting a {\mbox{non-commutative}} version would
  require the knowledge of an explicit solution to the Seiberg--Witten
 equations of non-Abelian gauge theory. On the other hand, describing
 an Abelian D-brane in terms of lower-dimensional ones in the language
 of matrix theory allows for identities involving  scalars, since the
 Abelian D-brane does not couple {\`a} la Myers~\cite{Myers} to transverse scalars. It
  should be possible to obtain from a {\mbox topological} identity an instanton equation including scalars, that would show up
 as infinite matrices.\\

\noindent{{{ \Large \bf Acknowledgements}}}\\ 

\noindent{I} thank R. Minasian for helpful discussions and mentoring
 on the subject of the present work, and {\mbox A. Tomasiello} for a critical
 reading of the manuscript.


\begin{thebibliography}{holomorphic}



 \bibitem {CH} C.-S. Chu and P.-M. Ho, \emph{Non-Commutative Open
     String and D-brane},
 {\mbox{Nucl. Phys.}} {\bf B550} (1999) 151-168,  {\ttfamily hep-th/9812219}. 


\bibitem {SW} N. Seiberg and E. Witten, \emph{String Theory and Non-Commutative Geometry}, JHEP {\bf 9909} (1999) 032, {\ttfamily hep-th/9908142}.


\bibitem {Cornalba} L. Cornalba, \emph{D-branes Physics and Non-Commutative
    Yang--Mills Theory}, {\mbox{Adv. Theor. Math. Phys.}} {\bf 4} (2000)
  271-281, {\ttfamily hep-th/9909081}.


\bibitem {DMS} S.R. Das, S. Mukhi and N.V. Suryanarayana, \emph{Derivative Corrections from {\mbox{Non-Commutativity}}}, JHEP {\bf 0108} (2001) 039, {\ttfamily hep-th/0106024}.



\bibitem{MMMS} M. Mari{\~n}o, R. Minasian, G. Moore and A. Strominger,
  \emph{Nonlinear Instantons from Supersymmetric $p$-branes}, JHEP {\bf 0001}
  (2000) 005, {\ttfamily hep-th/9911206}.





\bibitem  {Kapustin} A. Kapustin, \emph{Topological Strings on
    Non-Commutative Manifolds}, {\ttfamily hep-th/0310057}.




\bibitem  {Hitchin} N. Hitchin, \emph{Generalized Calabi--Yau
    manifolds},  Quart. J. Math. Oxford {\bf 54} (2003) 281-308, {\ttfamily math.DG/0209099}.




\bibitem  {Gualtieri} M. Gualtieri, \emph{Generalized Complex
    Geometry}, Ph.D. Thesis, Oxford, 2003, {\ttfamily math.DG/0401221}.





\bibitem{KapustinWS} A. Kapustin and Y. Li, \emph{Stability Conditions For Topological D-branes: A
  {\mbox{World-Sheet}} Approach}, {\ttfamily hep-th/0311101}.



\bibitem {MW} T. Mehen and M. Wise, \emph{Generalized *-Products,
    Wilson Lines and the Solution of the Seiberg--Witten Equations},
  JHEP {\bf 0012} (2000) 008, {\ttfamily hep-th/0010204}.



\bibitem  {Liu} H. Liu, \emph{$*$-Trek II: $*_n$ Operations, Open Wilson Lines and the Seiberg--Witten Map}, {\mbox{Nucl. Phys.}} {\bf B614} (2001) 305-329, {\ttfamily hep-th/0011125}.

  
\bibitem {OO} Y. Okawa and H. Ooguri, \emph{An Exact Solution to Seiberg--Witten
  {\mbox{Equation}} of {\mbox{Non-Commutative}} Gauge Theory}, Phys. Rev. {\bf
  D64} (2001) 046009, {\ttfamily hep-th/0104036}.



\bibitem {CSterms} S. Mukhi and N.V. Suryanarayana,
  \emph{Gauge-Invariant Couplings of {\mbox{Non-Commutative}} Branes to Ramond--Ramond Backgrounds}, JHEP {\bf{0105}} (2001) 023, {\ttfamily hep-th/0104045}.


\bibitem {MukhiSP} S. Mukhi, \emph{ Star Products from Commutative String Theory}, Pramana {\bf 58} (2002) 21-26, {\ttfamily hep-th/0108072}.



\bibitem {mezigue} P. Grange, \emph{Derivative Corrections from Boundary State Computations}, {\mbox{Nucl. Phys.}} {\bf
  B649} (2003) 297-311, {\ttfamily hep-th/0207211}.



\bibitem{Rey2} S.R. Das and S.-J. Rey, \emph{Open Wilson Lines in Non-Commutative Gauge Theory and Tomography of Holographic
      Dual Supergravity}, {\mbox{Nucl. Phys.}} {\bf B590} (2000) 453-470, {\ttfamily hep-th/0008042}.


\bibitem {Gross}  D.J. Gross, A. Hashimoto and N. Itzhaki, \emph{ Observables of Non-Commutative Gauge {\mbox{Theories}}}, Adv. Theor. Math. Phys. {\bf 4} (2000) 893-928, {\ttfamily hep-th/0008075}.


\bibitem{DT}  S.R. Das and S.P. Trivedi, \emph{ Supergravity Couplings
    to Non-Commutative Branes, Open Wilson Lines and Generalised Star Products}, JHEP {\bf 0102} (2001) 046, {\ttfamily hep-th/0011131}.


\bibitem {MSbeyond} S. Mukhi and N.V. Suryanarayana, \emph{Open-String
    Actions and Non-Commutativity {\mbox{Beyond}} the Large-B Limit},
  JHEP {\bf 0211} (2002) 002, {\ttfamily hep-th/0208203}.


  




\bibitem {beyond} P. Grange, \emph{Modified Star-Products Beyond the
    Large-$B$ Limit}, to appear in {\mbox{Phys. Lett.}} {\bf B}, {\ttfamily hep-th/0304059}.


\bibitem {variations} R. Minasian and A. Tomasiello, \emph{Variations on
    stability}, {\mbox{Nucl. Phys.}} {\bf{B631}} (2002) 43-65, {\ttfamily hep-th/0104041}.


\bibitem {Myers} R. Myers, \emph{Dielectric-Branes}, JHEP {\bf 9912} (1999)
  022, {\ttfamily hep-th/9910053}.
 

 
\end{thebibliography}
\end{document}